\documentclass[aps,prl,reprint,superscriptaddress]{revtex4-1}
\usepackage{amsmath}
\usepackage{amsfonts}
\usepackage{verbatim}
\usepackage{amsthm}
\usepackage{amssymb}
\usepackage{graphicx}
\usepackage{array}
\usepackage{enumitem}
\usepackage{float}
\usepackage{etoolbox}

\newcommand{\fig}{Fig.\ }
\newcommand{\NEO}{\bar{N}}

\begin{document}


\title{Invasion Rate versus Diversity in Population Dynamics with Catastrophes}

\author{A. Melka}
\affiliation{Department of Mathematics, Bar-Ilan University, Ramat Gan 52900, Israel}

\author{N. Dori}
\affiliation{Gonda Brain Research Center, Bar-Ilan University, Ramat Gan 52900, Israel}

\author{Y. Louzoun}
\email[Corresponding author: ]{louzouy@math.biu.ac.il}
\affiliation{Department of Mathematics, Bar-Ilan University, Ramat Gan 52900, Israel}
\affiliation{Gonda Brain Research Center, Bar-Ilan University, Ramat Gan 52900, Israel}

\date{\today}

\MessageBreak

\begin{abstract}
A key question in the current diversity crisis is how diversity has been maintained throughout evolution and how to preserve it. Modern coexistence theories suggest that a high invasion rate of rare new types is directly related to diversity. We show that adding almost any mechanism of catastrophes to a stochastic birth, death, and mutation process with limited carrying capacity induces a novel phase transition characterized by a positive invasion rate but a low diversity. In this phase, new types emerge and grow rapidly, but the resulting growth of very large types decreases diversity. This model also resolves two major drawbacks of neutral evolution models: their failure to explain balancing selection without resorting to fitness differences and the unrealistic time required for the creation of the observed large types. We test this model on a classical case of genetic polymorphism: the HLA locus.
\end{abstract}


\maketitle

The coexistence of numerous types, their diversity and their growth time have been longstanding subjects of interest and modeling \cite{hanski1994practical, smith1963population}. Complex models have been developed to explain coexistence, such as the Allee effect \cite{stephens1999allee, schreiber2003allee}, the rescue effect \cite{gotelli1991metapopulation}, depensation \cite{liermann2001depensation}, or density dependent growth \cite{fay2001positive, hassell1975density}. Ecologists such as Chesson \cite{chesson2000general, chesson2000mechanisms}, and more recently Ellner et al. \cite{ellner2019expanded} have merged many of those into a consistent model named the modern coexistence theory (MCT). They define the invasion rate as the probability for new types to grow in the presence of other competitive types. They suggest that mechanisms increasing the invasion rate contribute to coexistence and diversity. Diversity is commonly defined as the number of types over the total population \cite{, simpson1949measurement, nei1973analysis, hughes1997population}. 

While it would seem intuitive that diversity is indeed maintained through the emergence of new types, the mechanisms driving this emergence can in parallel promote the growth of very large types. Assuming that at equilibrium, the total population is fairly constant, and assuming a heavy-tailed distribution of type sizes, as is often observed and predicted \cite{ewens1972sampling, yuan2004structural}, those large types may occupy a macroscopic fraction of the total population and lower the diversity. We here use an extension of a classical birth, death, innovation (or mutation) neutral model (BDIM) including catastrophes (BDICM) \cite{dori2018family}. We demonstrate that the addition of catastrophes can simultaneously increase the invasion rate and decrease diversity, breaking the paradigm. A catastrophe represents a major deletion event where a type is fully or partly eliminated. In a constant resources environment, the total population size is balanced by equal average birth and death rates. However, when catastrophes are introduced, this balance is broken, and the average death rate of each type not hit by a catastrophe is less than its average birth rate. This results in net growth for each individual and exponential growth for all types not yet hit. Part of those births are mutations yielding new types of size 1. Their positive net growth is equivalent to a positive invasion rate. However, exponential growth also leads to very large types. The presence of those large types forces a smaller total number of types. We, therefore, observe a lower diversity despite a positive invasion rate.

We also show that BDICM resolves critical issues raised in classical neutral models  (e.g., the Moran model \cite{karlin1962genetics, kelly1977exact}). They indeed fail to explain the coexistence of several alleles at frequencies higher than prediction from genetic drift. The genetic drift is the change in the genetic composition of a population through random sampling of the current members dying or duplicating. This coexistence is commonly referred to as balancing selection \cite{key2014advantageous}. It is usually explained by frequency dependent selection (FDS), where the fitness of a genotype decreases as it becomes more common \cite{hedrick1983evidence, fitzpatrick2007maintaining}. While complex models have been proposed for balancing selection or negative FDS using fitness differences \cite{slater2015power, hedrick1983evidence, alter2017hla}, there is limited evidence for any of those. Classical models also fail to explain the emergence of large types and usually require an unrealistically long time for their growth. This problem, raised by Karev et al. \cite{karev2002birth, karev2003simple}, was partly solved in their nonlinear model, but it assumes that individual net growth rates are affected by their type size, which is not practical for genes or any other nonspatial characteristics. An interesting aspect of BDICM is that, in the presence of catastrophes, life expectancies become much lower and uniform across almost all type sizes since all types can get hit with equal probability. We show that this aspect addresses the issue of long expected growth time. It also replicates features of balancing selection in a purely neutral scenario since it removes the advantage of large types. 

\textit{The model.}\textemdash Formally, we expanded upon a neutral birth, death and, innovation process, by introducing large-scale events (catastrophes). We assume asexual reproduction. The number of individuals of each type is denoted by $k$ and the number of types of this size is denoted by $N_k$. The moments of the distribution are $m_j=\sum_k k^jN_k, \ j=0,1,2, $ where $j$ is the moment's order. Using this definition, $m_0$ is the number of types and $m_1$ is the number of individuals in the population.

\begin{figure}[h]
\begin{center}
\includegraphics[width=0.47\textwidth]{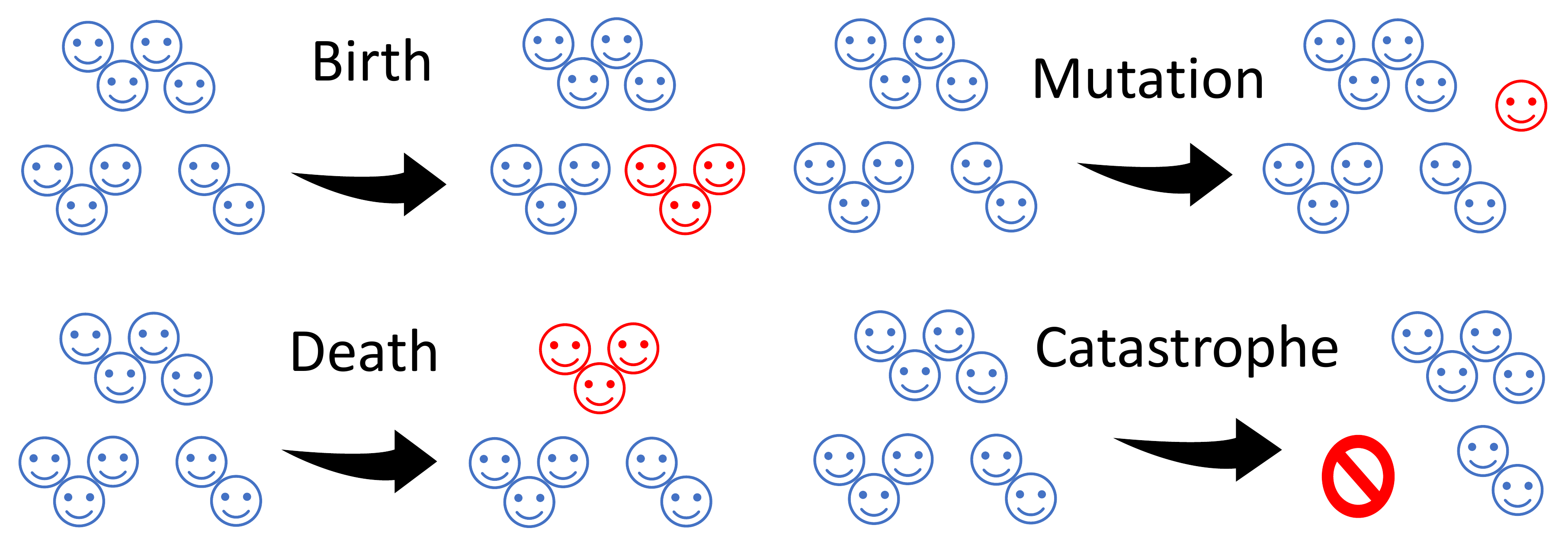}
\caption{Description of the model dynamics. Birth: the number of individuals in a given type increases by 1. Death: the number of individuals in a given type decreases by 1. Mutation: a constant fraction of all birth events leads to the creation of new types of size 1. Catastrophe: all individuals in a type are deleted.}
\label{fig:bdim}
\end{center}
\end{figure}

Birth events increase by 1 the number of individuals of a type and occur at a rate of $\alpha$ per individual. A constant fraction $\mu$ of all births leads through mutations to new types of size 1. Death events decrease by 1 the number of individuals of a type and occur at a rate of $\frac{m_1}{\NEO}$ per individual. $\NEO$ is the expected population size in equilibrium in the absence of catastrophes (see Eq. (2) in the Supplemental Material \cite{suppmat}). The model assumes that the death rate is proportional to the total population size to balance the total population as in the standard nutrient restricted logistic model. A catastrophe deletes all the individuals of a given type with a total rate of $\gamma$ (we further show that models with only partial deletion yields similar results). The probability that a type would be extinct in a catastrophe is not affected by its size. $\alpha$, $\gamma$, $\NEO$, and $\mu$ are free parameters. Without loss of generality, $\alpha$ can be set equal to 1 through a time rescaling (see \cite{dori2018family} and \fig \ref{fig:bdim} for a description of the model).

Time is discretized with time steps $\frac{1}{m_{1}}$. Within this short time interval, a type of size $k$ will either not change, grow to a type of size $k+1$, lose one of its members and become a type of size $k-1$ or disappear completely through a catastrophe or through a death if its size was 1. In equilibrium, total births and deaths are equal: $\frac{\alpha}{m_{1}} = \frac{1}{\NEO} + \frac{\gamma}{m_{0}}$. Since only nonmutation birth events increase the current type size, the probability to increase is $k \frac{\alpha (1-\mu)}{m_{1}}$. Similarly, the probability for a type to decrease by death is $\frac{k}{\NEO} = k \left( \frac{\alpha}{m_{1}} - \frac{\gamma}{m_{0}} \right)$ and by catastrophe is $\frac{\gamma}{m_{0}}$. Denoting by $T_{k}$ the average time to extinction for a type of initial size $k$, we obtain Eq. (\ref{eq:Tkk}). We solved this system with a matrix inversion (see the Supplemental Material \cite{suppmat} for derivations). To avoid restrictive assumptions on the type size distribution (as in \cite{dori2018family}), we used numerical estimates for $m_0$ and $m_1$ obtained from simulations reaching steady state, for each value of $\gamma$.

\begin{equation}
\left\{
\begin{aligned}
T_{1} & = 1 + \frac{\alpha (1 - \mu)}{m_{1}}T_{2} + \left[1 - \frac{\alpha (1 - \mu)}{m_{1}} - \left( \frac{\alpha}{m_{1}} - \frac{\gamma}{m_{0}} \right) \right]T_{1} \\
T_{k} & = 1 + k \frac{\alpha (1 - \mu)}{m_{1}}T_{k+1} + k \left(\frac{\alpha}{m_{1}} - \frac{\gamma}{m_{0}} \right) T_{k-1} \\
& + \left[1 - k\frac{\alpha (1 - \mu)}{m_{1}} - k \left(\frac{\alpha}{m_{1}} - \frac{\gamma}{m_{0}} \right) - \frac{\gamma}{m_{0}} \right]T_{k} .
\end{aligned}
\right.
\label{eq:Tkk}
\end{equation}

We derived similar results using a Galton-Watson (GW) process. In a typical GW process, a population is assumed to have an initial size of 1. At each time step, each member can be replaced by $i$ identical descendants. Here, we set $i \in {0,1,2}$. Death corresponds to $i = 0$, no event to $i = 1$, and birth to $i = 2$. We denote by $p_{i}$ the probability of getting $i$ descendants. The time to extinction in a regular GW process, namely when the population reaches a size of 0, is obtained through the probability generating function $G(x) = p_{0} + p_{1} x + p_{2} x^2$. Denoting by $d_{t}$ the probability for this type to be extinct at time $t$, the GW model states that $d_{t} = G(d_{t-1})$. If the initial size is $k$ instead of 1, then the probability to be extinct at time $t$ is $d_{t}^{k}$ since all $k$ members behave like independent types of size 1, ignoring the possible interactions through the logistic term and assuming a large total population. Since the GW process only describes the events of regular birth or death and does not account for the catastrophes, the survival probability must be multiplied by the probability that no catastrophe occurred. Denoting by $\tilde{d}_{t,k} = 1 - (1 - d_{t}^{k}) e^{\frac{-\gamma t}{m_{0}}}$, the catastrophe corrected probability of extinction, one gets:

\begin{equation}
\begin{aligned}
p_{0} = \frac{\alpha}{m_{1}} - \frac{\gamma}{m_{0}}, \> p_{1} &= 1 - p_{0} - p_{2}, \> p_{2} = \frac{\alpha(1 - \mu)}{m_{1}}, \\
T_{k} &= \sum_{t}t(\tilde{d}_{t,k} - \tilde{d}_{t-1,k}) .
\end{aligned}
\end{equation} 

As expected, the GW process leads to the same times to extinction as in the analytical method. Finally, we show the accuracy of our two approaches by testing them against simulations (lower plots in \fig \ref{fig:TimeExt}), with similar results (see the Supplemental Material \cite{suppmat} for methodologies). 

In classical BDIM (i.e., no catastrophe) \cite{karev2002birth, karlin1962genetics}, there is a natural extinction survival transition at a zero net growth rate, and the times to extinction increase with respect to the initial type size (green lines in upper right plot of \fig \ref {fig:TimeExt}). However, in the BDICM, two other transitions appear. If $\gamma$ is higher than $\mu$, a transition to extinction occurs since fewer new types are created by mutations than the ones destroyed. As $\gamma$ approaches $\mu$, another intermediate transition is observed, where the times to extinction not only decrease but also become uniform across all initial type sizes (upper left plot in \fig \ref {fig:TimeExt}) resulting in a change in the balance between small and large types. As will be explained in the next section, this transition is characterized by a positive invasion rate since large types can now disappear just as fast as small ones and faster than for $\gamma=0$, creating room for new types to appear and grow. It is also characterized by a lower diversity since types can grow rapidly, before getting hit, to attain large sizes (longer range in the blue lines in upper right plot of \fig \ref {fig:TimeExt}) and hold a large fraction of the total population.

\begin{figure}[h]
\begin{center}
\includegraphics[width=0.48\textwidth]{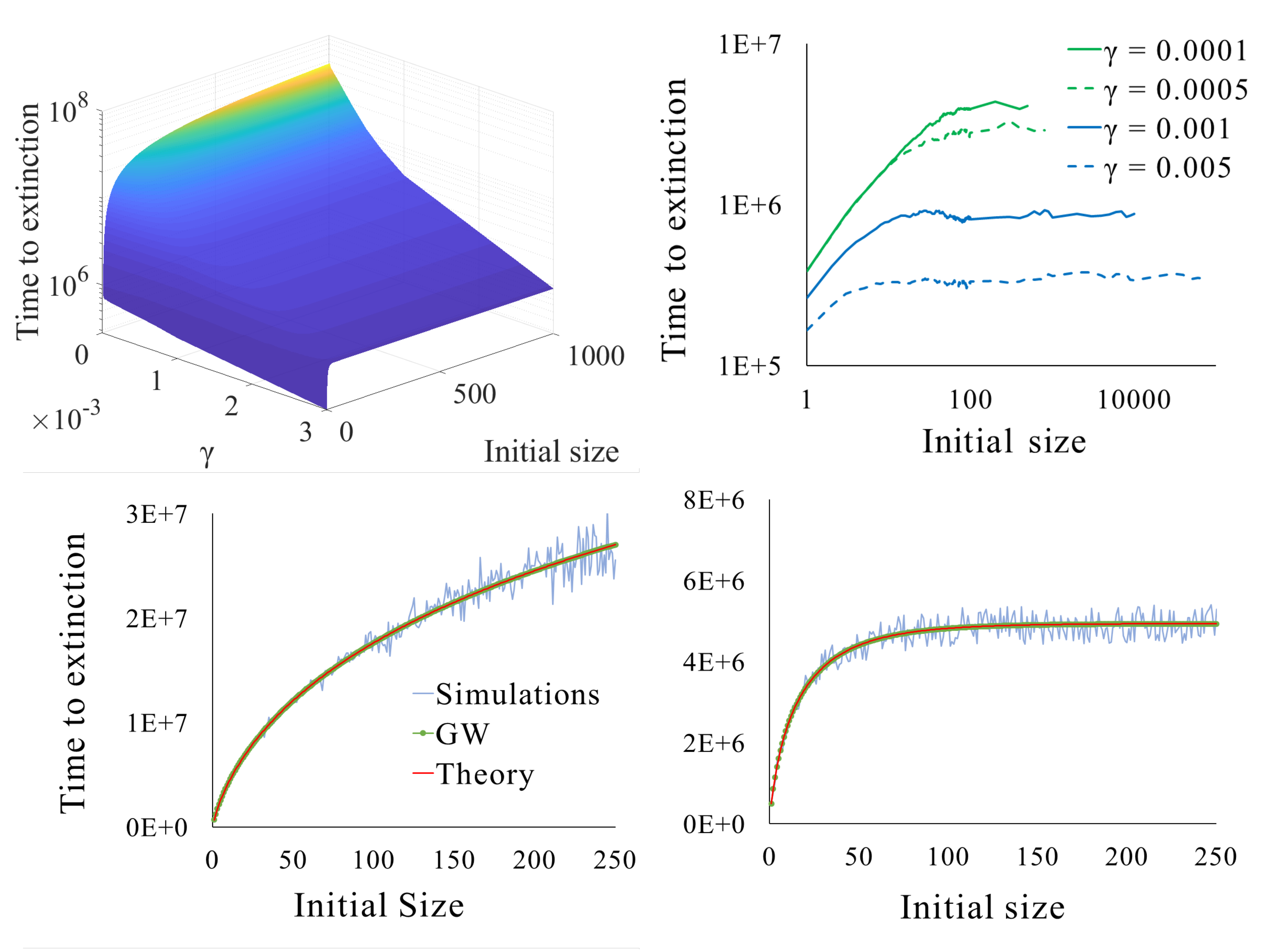}
\caption{Upper left plot: times to extinction with respect to the catastrophe rate $\gamma$ and the initial type size. Upper right plot: times to extinction for different values of $\gamma$. Green curves correspond to the ``negative invasion rate" phase where large types live longer but never reach a high size. Blue curves correspond to the ``positive invasion rate" phase where catastrophes level the lifetime expectancy but types can reach higher sizes. Lower plots: comparison between simulations, GW process and analytical expected times to extinction. The left plot is for $\gamma = 0$ and the right plot for $\gamma = 0.001$. Without catastrophe, times to extinction grow with type size and can reach high values. As $\gamma$ increases, times to extinction are reduced and become uniform across type sizes for sufficiently large types.}
\label{fig:TimeExt}
\end{center}
\end{figure}

\textit{Invasion rate vs. diversity.}\textemdash Using the GW model, one can compute the average growth after one step as:

\begin{equation}
\begin{aligned}
\mathbb{E}[\text{growth after one step}] &= \ln \left[ \frac{1}{k} \left. \frac{d[G(x)]^{k}}{dx} \right|_{x=1} \right] \\
&= \ln{(1 - p_{0} + p_{2})} .
\end{aligned}
\end{equation} 

\begin{figure}[h]
\begin{center}
\includegraphics[width=0.48\textwidth]{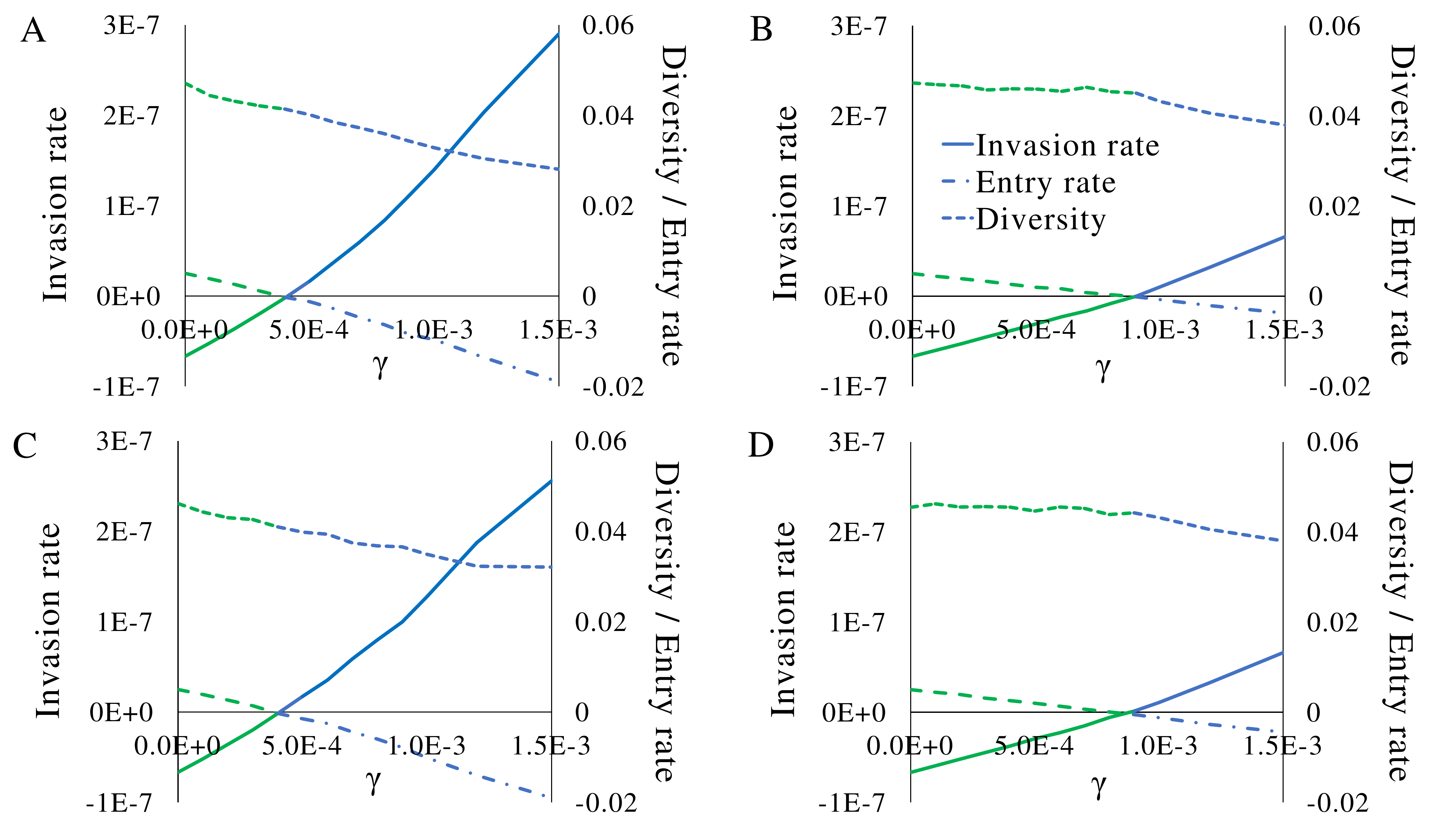}
\caption{On each plot the right scale represents the diversity as defined as $m_{0} / m_{1}$ and the entry rate as defined as the number of mutations minus the deaths by catastrophe. The left scale represents the invasion rate as the log of the average growth rate of types after one step. (a) Complete deletion for each catastrophe as baseline model. (b) Deletion of a fixed fraction of a type in each catastrophe. (c) Catastrophe rate proportional to the population size. (d) Deletion of a random fraction of a type. We observe in all models that above a given catastrophe rate, the invasion rate becomes positive while diversity decreases.}
\label{fig:DivInvEntry}
\end{center}
\end{figure}

Note that this growth rate is independent of the initial type size $k$ and its sign is entirely determined by the sign of $p_{2} - p_{0}$. Without catastrophes, on average, existing type sizes would consistently decrease (since new types are created, and the total birth and death rate must equilibrate). This decrease leads them to the 0 absorbing state. The presence of catastrophes ensures that, in equilibrium, a transition occurs for $\gamma = \frac{\alpha \mu m_{0}}{m_{1}}$, where the average growth rate of all types becomes positive until their annihilation by a catastrophe as shown in \fig \ref{fig:DivInvEntry}. For this same $\gamma$ value, we also observe from the simulations that the slope of the diversity ($m_{0} / m_{1}$) becomes more negative and the entry rate becomes negative, which confirms the results obtained with the GW model. The entry rate is defined as the difference between the number of births by mutation and the number of deaths by catastrophe. Since the equilibrium ensures that regular births plus mutations equal natural deaths plus catastrophes, the entry rate is opposite to the growth rate. 

This transition, although closely related, is different from the one observed in \cite{ dori2018family}. The transition described there is between ``low variance" and ``high variance" phases for $\gamma = \frac{2 \alpha \mu m_{0}}{m_{1}}$. Our transition appears at a lower $\gamma$, and the diversity starts to decrease before the second moment of the type size distribution diverges. 

So far, the catastrophe rate was neither a function of the population size nor the number of types. When a catastrophe event occurs, a type is randomly chosen to be eliminated with equal probability over all types. To assess the robustness of our results, we compared other catastrophe models: two in which, for each catastrophe, only a fraction (fixed or random) of the population of one type is destroyed, and one where the catastrophe rate is proportional to $m_1$ like the regular death rate (see the Supplemental Material \cite{suppmat}). As shown in \fig \ref {fig:DivInvEntry}, as $\gamma$ increases, the fraction of deaths induced by catastrophes becomes large enough and the transition emerges.

To summarize, in BDICM, in the ``positive invasion rate" phase, although the lifespan of each type is on average low since they can be deleted via catastrophes, the growth rate is positive across all types. Therefore, several new types emerge and rapidly invade the system. At the same time, surviving types grow exponentially and become so large that the diversity decreases, breaking the paradigm that a positive invasion rate implies higher diversity. 

\textit{Comparison to FDS Models.}\textemdash This ``positive invasion rate" phase is similar to balancing selection in that it promotes the growth of rare types. Moreover, in BDIM, small types have a disadvantage since they have a higher probability to reach the 0 absorbing state. However, when death is dominated by catastrophes, large types also have a high probability to reach the absorbing state. This removes their advantage over small types and gives the later a higher relative chance to grow. Other models were developed to explain balancing selection. For instance, Karev et al. \cite{karev2003simple} proposed a nonlinear BDIM, which corresponds to FDS. In their linear BDIM (without FDS), birth and death rates are equal and proportional to the type size. This model does not explain balancing selection and it requires extremely long times to reach equilibrium. Those issues are resolved in their nonlinear FDS model. However, this model implies the unrealistic assumption that individuals die or give birth knowing the size of their type. Such an assumption might occur if each type had a distinct resource, but it is highly improbable in a model with hundreds of types.

\begin{figure}[h]
\begin{center}
\includegraphics[width=0.48\textwidth]{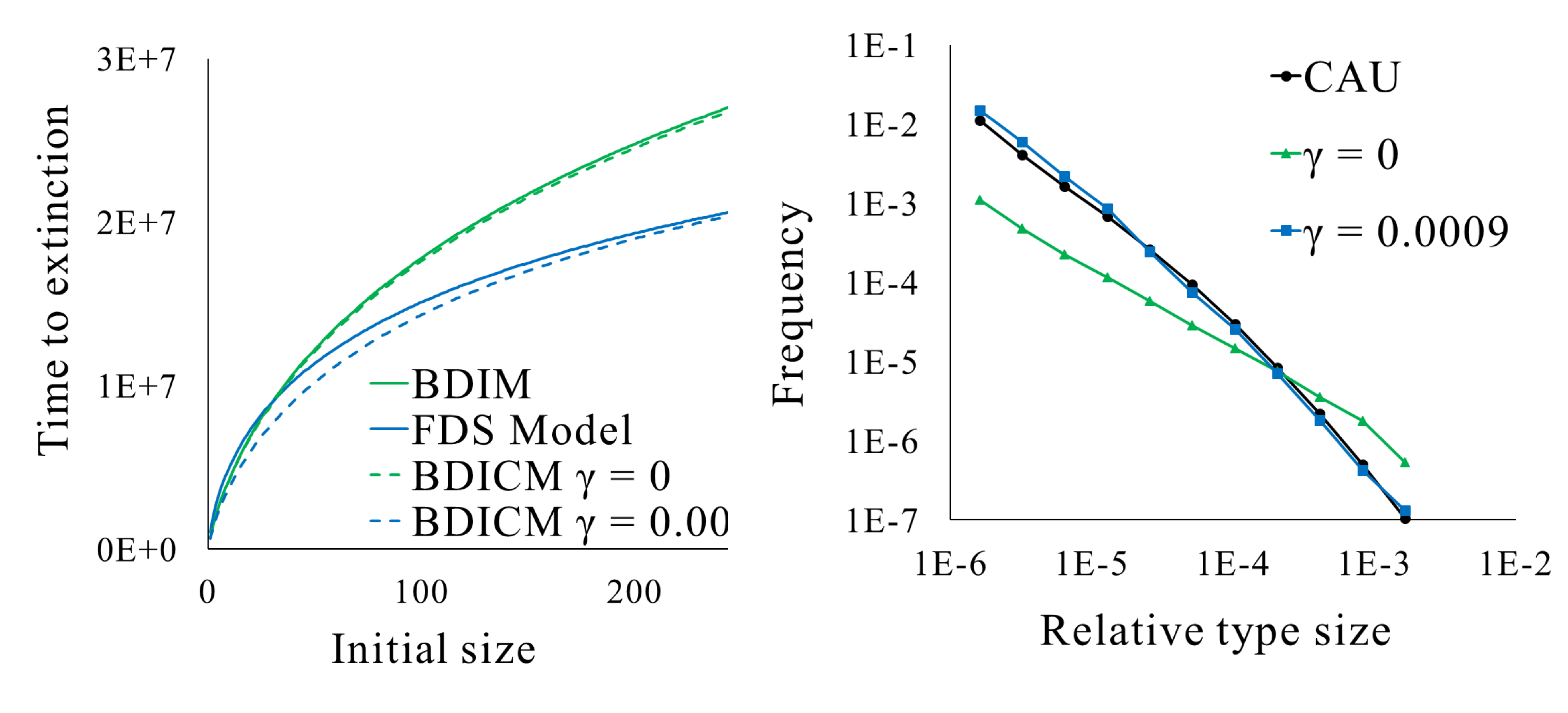}
\caption{Left plot: simulated and analytical times to extinction computed with the modified model with FDS introduced in the form of a constant birth rate $a$ and a reduced relative birth rate $\alpha$. For $a = 0$, the curves fit with our model without catastrophe. When adding balancing selection, we observe that the behavior is similar to adding catastrophes. Hence, the catastrophe model replicates balancing selection. Right plot: allele frequencies. The CAU curves represent the frequencies of HLA alleles for the Caucasian population. We compared it with simulations with and without catastrophes. We observe that the presence of catastrophes reproduces the frequencies of the alleles and could explain their distribution.}
\label{fig:karevdist}
\end{center}
\end{figure}

The BDICM with $\gamma=0$ reproduces the same times to extinction as the linear model \cite{karev2003simple}. In the nonlinear model, a positive fixed birth rate $a$ (type size independent) is assumed for each type and the individual birth rate $\alpha$ is set to be lower to achieve equilibrium $\frac{\alpha (k + a)}{m_{1}} = \frac{k}{\NEO}$ (see the Supplemental Material \cite{suppmat}). The times to extinction with FDS are equivalent to the ones obtained for a weak catastrophe rate, suggesting a similarity between classical negative FDS and the BDICM (\fig \ref{fig:karevdist}). 

\textit{Populations in HLA.}\textemdash One of the systems where the contrast between diversity and a positive invasion rate is the clearest is the MHC locus \cite{de1982structure}. MHC proteins present self and foreign peptides to immune system cells. The genes coding for these proteins (denoted HLA in humans) are by far the most polymorphic in the human genome with thousands of alleles of each of the main classical HLA genes \cite{slater2015power, zachary1996frequencies}. The main arguments currently used to explain the origins of this extreme genetic diversity in a specific locus are different aspects of balancing selection where new types have an advantage over existing types \cite{hedrick1983evidence}. However, recent evidence suggests that the HLA haplotype distribution is biased toward excessive large types \cite{alter2017hla, lobkovsky2019multiplicative, van2009new, sommer2005importance}.

To test whether BDICM is consistent with the HLA frequencies, we computed the HLA-A frequency distribution in different populations in the United States, and the catastrophe rate producing the best fit for this distribution. A catastrophe rate of 0 cannot explain the observed distribution. However, a low catastrophe rate can produce an excellent fit (see \fig \ref{fig:karevdist} for the Caucasian population). This is not evidence that catastrophes are the mechanism explaining the distribution, but it shows their plausibility. In this context, catastrophes would represent a destructive allele, that, in some conditions, can kill all individuals that carry it. The distributions for other alleles and populations are similar \cite{slater2015power, alter2017hla, lobkovsky2019multiplicative}.

\textit{Conclusions.}\textemdash Catastrophes partly or fully erasing populations are frequent in many domains ranging from ecology to market dynamics \cite{hsieh2005distinguishing, knight1999corporate, thomann2013impact}. In the population dynamics context, such events can be the eradication of local populations through weather events or diseases. In the genetic context, this can happen if a gene induces susceptibility to a certain pathogen. Introducing even a small amount of catastrophes to a BDIM inherently changes the dynamics. Specifically, there is a transition to an ``invasion" regime, where the lifespan of types is mainly determined by catastrophes and not by the diffusion to the 0 absorbing state. 

The intuition is simple. In the absence of catastrophes, births and deaths must be balanced leading to an Ewens-like type-size distribution \cite{ewens1972sampling}. However, in the presence of catastrophes, the average birth rate of each type can be higher than its death rate, with the total population balanced by catastrophes. This results in a positive growth rate for all type sizes and a positive invasion rate. In such a case, the sizes of types are determined by the combination of exponential growth and geometric distribution of time to catastrophe, leading to a skewing of the distribution toward large types. These large types can exist in parallel with a high invasion rate since the lifespan of each type is very low. 

This model demonstrates that the paradigm in which the invasion rate induces diversity does not hold. Moreover, it shows that, in order to maintain diversity, the focus should be on limiting the size of large types instead of boosting the invasion rate. The presence of catastrophes can explain balancing selection without resorting to fitness differences or nonlinear models. It also explains the emergence and growth of large types within short times. Since this transition is only the result of a change in the balance between the total birth and death events, it is not affected by the details of the catastrophes. Similarly, Wilcox et al. \cite{wilcox2003effect} showed that density dependent catastrophes do not change the conclusions of their model. 

\bibliographystyle{apsrev4-1}

\end{document}